\pdfoutput=1

\documentclass[runningheads,a4paper,english]{llncs}[2022/01/12]

\usepackage[utf8]{inputenc}

\usepackage{upquote}

\usepackage[ngerman,main=english]{babel}
\addto\extrasenglish{\languageshorthands{ngerman}\useshorthands{"}}
\usepackage{regexpatch}
\makeatletter
\edef\switcht@albion{%
  \relax\unexpanded\expandafter{\switcht@albion}%
}
\xpatchcmd*{\switcht@albion}{ \def}{\def}{}{}
\xpatchcmd{\switcht@albion}{\relax}{}{}{}
\edef\switcht@deutsch{%
  \relax\unexpanded\expandafter{\switcht@deutsch}%
}
\xpatchcmd*{\switcht@deutsch}{ \def}{\def}{}{}
\xpatchcmd{\switcht@deutsch}{\relax}{}{}{}
\edef\switcht@francais{%
  \relax\unexpanded\expandafter{\switcht@francais}%
}
\xpatchcmd*{\switcht@francais}{ \def}{\def}{}{}
\xpatchcmd{\switcht@francais}{\relax}{}{}{}
\makeatother

\usepackage[hyphens]{url}

\makeatletter
\g@addto@macro{\UrlBreaks}{\UrlOrds}
\makeatother

\usepackage[%
    rm={oldstyle=false,proportional=true},%
    sf={oldstyle=false,proportional=true},%
    tt={oldstyle=false,proportional=true,variable=false},%
    qt=false%
]{cfr-lm}

\usepackage[T1]{fontenc}

\usepackage[
  babel=true, %
  expansion=alltext,
  protrusion=alltext-nott, %
  final %
]{microtype}

\DisableLigatures{encoding = T1, family = tt* }

\usepackage[pdftex]{graphicx}

\usepackage{diagbox}

\usepackage[table]{xcolor}

\usepackage[autostyle=true]{csquotes}

\defineshorthand{"`}{\openautoquote}
\defineshorthand{"'}{\closeautoquote}

\usepackage{booktabs}

\usepackage{paralist}

\usepackage[rflt]{floatflt}

\usepackage{stfloats}
\fnbelowfloat

\usepackage[group-minimum-digits=4,per-mode=fraction]{siunitx}
\addto\extrasgerman{\sisetup{locale = DE}}

\usepackage{hyperref}

\hypersetup{
  hidelinks,
  colorlinks=true,
  allcolors=black,
  pdfstartview=Fit,
  breaklinks=true
}

\usepackage[all]{hypcap}

\usepackage{mindflow}

\usepackage{amsmath}
\usepackage[capitalise,nameinlink]{cleveref}

\crefname{section}{Sect.}{Sect.}
\Crefname{section}{Section}{Sections}
\crefname{listing}{List.}{List.}
\crefname{listing}{Listing}{Listings}
\Crefname{listing}{Listing}{Listings}
\crefname{lstlisting}{Listing}{Listings}
\Crefname{lstlisting}{Listing}{Listings}

\DeclareFontFamily{U}{MnSymbolC}{}
\DeclareSymbolFont{MnSyC}{U}{MnSymbolC}{m}{n}
\DeclareFontShape{U}{MnSymbolC}{m}{n}{
  <-6>    MnSymbolC5
  <6-7>   MnSymbolC6
  <7-8>   MnSymbolC7
  <8-9>   MnSymbolC8
  <9-10>  MnSymbolC9
  <10-12> MnSymbolC10
  <12->   MnSymbolC12%
}{}
\DeclareMathSymbol{\powerset}{\mathord}{MnSyC}{180}

\usepackage{xspace}
\newcommand*{\eg}{e.\,g.,\ }

\makeatletter
\newcommand{\hydash}{\penalty\@M-\hskip\z@skip}
\makeatother

\input glyphtounicode
\usepackage[T1]{fontenc}
\usepackage[utf8]{inputenc}
\graphicspath{ {plots} }

\usepackage{subcaption}
\usepackage{tikz}
\usetikzlibrary{external}
\usepackage{epstopdf}
\usepackage{tikzscale}

\usepackage{pgfplots}
\DeclareUnicodeCharacter{2212}{−}
\usepgfplotslibrary{groupplots,dateplot}
\usetikzlibrary{patterns,shapes.arrows}
\pgfplotsset{compat=newest}

\usepackage{csquotes}
\MakeOuterQuote{"}

\usepackage[disable]{todonotes}

\usepackage{listings}

\newcommand\lstcolor[1]{\color{#1!70!gray}}

\definecolor{lstnumber}{rgb}{0.3,0.3,0.3}
\definecolor{lstbackgroundcolor}{rgb}{0.96,0.96,0.96}

\colorlet{punct}{red!60!black}
\definecolor{background}{HTML}{EEEEEE}
\definecolor{delim}{RGB}{20,105,176}
\colorlet{numb}{magenta!60!black}

\lstdefinelanguage{json}{
    showstringspaces=false,
    breaklines=true,
    literate=
     *{0}{{{\color{numb}0}}}{1}
      {1}{{{\color{numb}1}}}{1}
      {2}{{{\color{numb}2}}}{1}
      {3}{{{\color{numb}3}}}{1}
      {4}{{{\color{numb}4}}}{1}
      {5}{{{\color{numb}5}}}{1}
      {6}{{{\color{numb}6}}}{1}
      {7}{{{\color{numb}7}}}{1}
      {8}{{{\color{numb}8}}}{1}
      {9}{{{\color{numb}9}}}{1}
      {:}{{{\color{punct}{:}}}}{1}
      {,}{{{\color{punct}{,}}}}{1}
      {\{}{{{\color{delim}{\{}}}}{1}
      {\}}{{{\color{delim}{\}}}}}{1}
      {[}{{{\color{delim}{[}}}}{1}
      {]}{{{\color{delim}{]}}}}{1},
}

\lstdefinelanguage{diff}{
  sensitive=true,
  morecomment=[f][\lstcolor{gray}][0]{diff},
  morecomment=[f][\lstcolor{gray}][0]{index},
  morecomment=[f][\lstcolor{blue}][0]{@@},
  morecomment=[f][\lstcolor{magenta}][0]{***},
  morecomment=[f][\lstcolor{violet}][0]{!},
  morecomment=[f][\lstcolor{red!60!black}][0]-,
  morecomment=[f][\lstcolor{green!60!black}][0]+,
  morecomment=[f][\lstcolor{red!60!black}][0]{---},
  morecomment=[f][\lstcolor{green!60!black}][0]{+++},
  morecomment=[f][\lstcolor{gray}][0]{Binary},
  morecomment=[f][\lstcolor{gray}][0]{Only},
  morecomment=[f][\lstcolor{gray}][0]{old},
  morecomment=[f][\lstcolor{gray}][0]{new},
  morecomment=[f][\lstcolor{gray}][0]{rename},
  morecomment=[f][\lstcolor{gray}][0]{similarity},
  morecomment=[f][\lstcolor{gray}][0]{deleted},
  morecomment=[f][\lstcolor{magenta}][0]{***************},
  morecomment=[f][\lstcolor{red!60!black}][0]<,
  morecomment=[f][\lstcolor{green!60!black}][0]>,
  morecomment=[f][\lstcolor{blue}][0]{0},
  morecomment=[f][\lstcolor{blue}][0]{1},
  morecomment=[f][\lstcolor{blue}][0]{2},
  morecomment=[f][\lstcolor{blue}][0]{3},
  morecomment=[f][\lstcolor{blue}][0]{4},
  morecomment=[f][\lstcolor{blue}][0]{5},
  morecomment=[f][\lstcolor{blue}][0]{6},
  morecomment=[f][\lstcolor{blue}][0]{7},
  morecomment=[f][\lstcolor{blue}][0]{8},
  morecomment=[f][\lstcolor{blue}][0]{9},
}[comments]

\lstdefinestyle{mystyle}{
    backgroundcolor=\color{lstbackgroundcolor},
    numberstyle=\tiny\color{lstnumber},
    basicstyle=\ttfamily\scriptsize,
    frame=single,
    framerule=0pt,
    framesep=2pt,
    xleftmargin=2pt,
    xrightmargin=2pt,
    numbersep=7pt,
    breakatwhitespace=false,
    breaklines=true,
    captionpos=b,
    keepspaces=true,
    numbers=left,
    showspaces=false,
    showstringspaces=false,
    showtabs=false,
    tabsize=4
}

\usepackage[backend=biber, style=lncs]{biblatex}
\bibliography{paper}

\usepackage[acronym,toc]{glossaries}
\newacronym{tapasco}{TaPaSCo}{Task Parallel System Composer}
\newacronym{pe}{PE}{\acrshort{tapasco} Processing Element}
\newacronym{dse}{DSE}{Design Space Exploration}

\newacronym{nist}{NIST}{National Institute of Standards and Technology}
\newacronym{pqc}{PQC}{Post-Quantum Cryptography}
\newacronym{kem}{KEM}{Key Encapsulation Mechanism}
\newacronym{dsa}{DSA}{Digital Signature Algorithm}
\newacronym{kat}{KAT}{Known Answer Test}
\newacronym{rng}{RNG}{Random Number Generator}
\newacronym{aes}{AES}{Advanced Encryption Standard}
\newacronym{sha-3}{SHA-3}{Secure Hash Algorithm 3}
\newacronym{svp}{SVP}{Shortest Vector Problem}
\newacronym{lwe}{LWE}{Learning With Errors}
\newacronym{sis}{SIS}{Short Integer Solution}
\newacronym{ntt}{NTT}{Number Theoretic Transform}
\newacronym{dft}{DFT}{Discrete Fourier Transform}
\newacronym{lfsr}{LFSR}{Linear-feedback Shift Register}

\newacronym{ind-cpa}{IND-CPA}{Indistinguishability under Chosen-Plaintext-Attack}
\newacronym{ind-cca}{IND-CCA}{Indistinguishability under Chosen-Ciphertext-Attack}
\newacronym{suf-cma}{SUF-CMA}{Strong Existential Unforgeability under Chosen-Message-Attack}

\newacronym{fpga}{FPGA}{Field-programmable Gate Array}
\newacronym{asic}{ASIC}{Application-specific Integrated Circuit}
\newacronym{hls}{HLS}{High-level Synthesis}
\newacronym{hdl}{HDL}{Hardware Description Language}
\newacronym{dma}{DMA}{Direct Memory Access}
\newacronym{mmu}{MMU}{Memory Management Unit}
\newacronym{axi4}{AXI4}{Advanced Extensible Interface 4}

\newacronym{lut}{LUT}{Lookup Table}
\newacronym{ff}{FF}{Flip Flop}
\newacronym{bram}{BRAM}{Block Random Access Memory}
\newacronym{dsp}{DSP}{Digital Signal Processor}

\newacronym{vc}{VC709}{Xilinx Virtex-7 VC709 \acrshort{fpga}}
\newacronym{au}{U280}{Xilinx Alveo U280 \acrshort{fpga}}

\newacronym{iot}{IoT}{Internet of Things}

\newacronym{lap}{LAP}{Latency-area Product}

\newacronym{cpu}{CPU}{Central Processing Unit}
\newacronym{avx2}{AVX2}{Advanced Vector Extensions 2}
\newacronym{risc}{RISC}{Reduced Instruction Set Computer}
\newacronym{isa}{ISA}{Instruction Set Architecture}
\newacronym{pcie}{PCIe}{Peripheral Component Interconnect Express}
\newacronym{ila}{ILA}{Integrated Logic Analyzer}

\newacronym{ci}{CI}{Continuous Integration}
\newacronym{json}{JSON}{JavaScript Object Notation}

\usepackage{tabularx}   %
\usepackage{longtable}  %
\usepackage{booktabs}   %
\usepackage{lscape}     %

\usepackage{etoolbox}
\robustify\itshape
\usepackage{xspace}
\newcommand*{\tapasco}{\acrshort{tapasco}\xspace}
\newcommand*{\pqc}{\acrshort{pqc}\xspace}
\newcommand*{\nist}{\acrshort{nist}\xspace}
\newcommand*{\kat}{\acrshort{kat}\xspace}
\newcommand*{\npp}{\acrshort{nist} \acrshort{pqc} project\xspace}
\newcommand*{\fpga}{\acrshort{fpga}\xspace}
\newcommand*{\fpgas}{\acrshortpl{fpga}\xspace}
\newcommand*{\hls}{\gls{hls}\xspace}
\newcommand*{\pe}{\acrshort{pe}\xspace}
\DeclareRobustCommand{\pes}{\acrshortpl{pe}\xspace}
\newcommand*{\dse}{\acrshort{dse}\xspace}
\newcommand*{\luts}{\acrshortpl{lut}\xspace}
\newcommand*{\ffs}{\acrshortpl{ff}\xspace}
\newcommand*{\brams}{\acrshortpl{bram}\xspace}
\newcommand*{\dsps}{\acrshortpl{dsp}\xspace}
\newcommand*{\riscv}{\acrshort{risc}-V\xspace}

\newcommand*{\vc}{\acrshort{vc}\xspace}
\newcommand*{\au}{\acrshort{au}\xspace}

\usepackage{color}

\begin{document}
\title{PQC-HA: A Framework for Prototyping and In-Hardware Evaluation of Post-Quantum Cryptography Hardware Accelerators
\thanks{
This research was funded by the German Federal Ministry for Education and Research (BMBF) with the funding ID 01IS17050.
This research work was supported by the National Research Center for Applied Cybersecurity ATHENE. ATHENE is funded jointly by the
German Federal Ministry of Education and Research and the Hessian Ministry of Higher Education, Research and the Arts
}
}
\titlerunning{PQC-HA: Prototyping and Evaluating PQC Hardware Accelerators}
\author{
Richard Sattel\orcidID{0009-0003-1060-3462}
\and
Christoph Spang\orcidID{0000-0003-1606-4474}
\and
Carsten Heinz\orcidID{0000-0001-5927-4426}
\and
Andreas Koch\orcidID{0000-0002-1164-3082}
}
\authorrunning{R. Sattel et al.}
\institute{Embedded Systems and Applications Group, TU Darmstadt, Germany \\
\email{\{sattel, spang, heinz, koch\}@esa.tu-darmstadt.de}\\
}
\maketitle%
\begin{abstract}
In the third round of the \nist \acrlong{pqc} standardization project, the focus is on optimizing software and hardware implementations of candidate schemes. The winning schemes are CRYSTALS Kyber and CRYSTALS Dilithium, which serve as a \acrfull{kem} and \acrfull{dsa}, respectively.
This study utilizes the \tapasco open-source framework to create hardware building blocks for both schemes using \hls from minimally modified ANSI C software reference implementations across all security levels. Additionally, a generic \tapasco host runtime application is developed in Rust to verify their functionality through the standard \nist interface, utilizing the corresponding \acrlong{kat} mechanism on actual hardware.
Building on this foundation, the communication overhead for \tapasco hardware accelerators on \acrshort{pcie}-connected FPGA devices is evaluated and compared with previous work and optimized \acrshort{avx2} software reference implementations. The results demonstrate the feasibility of verifying and evaluating the performance of Post-Quantum Cryptography accelerators on real hardware using \tapasco. Furthermore, the off-chip accelerator communication overhead of the \nist standard interface is measured, which, on its own, outweighs the execution wall clock time of the optimized software reference implementation of Kyber at Security Level 1.

\keywords{
Post-Quantum Cryptography
\and NIST
\and CRYSTALS Kyber
\and CRYSTALS Dilithium
\and FPGA
\and Hardware Accelerator
\and High-level Synthesis
\and TaPaSCo
}
\end{abstract}
\section{Introduction}
\label{sec:introduction}

From the perspective of recent progress in the field of quantum computing,
the \gls{nist} is setting out to standardize a set of algorithms for \gls{pqc}
to replace the majority of public-key cryptography in use today that will become
insecure with the advent of large quantum computers leaving most %
communication unprotected. The third round of this competition
elects CRYSTALS Kyber and CRYSTALS Dilithium as winners, for a \gls{kem} and \gls{dsa} respectively, with
a focus on
optimized software and hardware implementations.
The \acrfull{tapasco} open-source
framework provides essential components to aid in the design process of these
hardware accelerators, such as a generic hardware architecture in addition to a
Linux kernel driver and userspace runtime library, which complement one another
to schedule jobs on \acrfullpl{pe}\cite{heinz_tapasco_2021}.
In this work, we present the basis for the integration of \gls{pqc} building
blocks into the \tapasco framework. %
We use \acrfull{hls} to transform the top-level functions of CRYSTALS Kyber and CRYSTALS Dilithium into
\pes, which are hardware kernels containing logic to solve
a specific task.
We synthesize \pes of all \gls{nist} security levels for \textit{Encapsulation}
and \textit{Decapsulation} for Kyber as well as \textit{Sign} and \textit{Verify} for
Dilithium.
The \pes receive their parameters from a host runtime application
implemented in Rust using the standard \gls{nist} interface, which is a set of
functions that each candidate of the \npp needs to implement. It is used to test
the functional correctness of the \pes via the \gls{nist}
\gls{kat} cases, which are in turn verified to be correct with upstream
repository hashes, to make sure
they generate the same results as the correct upstream reference software
implementation. To enable verification of algorithms that depend on random input data
deterministically and at the same time avoid using insecure \gls{rng} units on
the \acrfull{fpga},
we extend the \gls{nist} interface to supply the
\pes with pre-initialized random buffers from the host.
The runtime provides generic functions for \gls{kem} and
\gls{dsa} algorithms and defines an extendable interface, which facilitates the
integration of further algorithms of the \npp as long as they adhere
to the \gls{nist} mandatory interface, optionally including the random buffer extension.
The rest of this work is organized as follows.
In \cref{sec:related_work}, we discuss related work for \hls and \gls{hdl} implementations, which we will later compare to our implementations.
In \cref{sec:approach}, we define our approach.
We describe
how we use \hls to create \tapasco \pes
from the current software ANSI C reference implementation using the same
standard interface for parameters.
In \cref{sec:evaluation}, we present the results of our designs, compare
them to related work and analyze the communication overhead of the chosen
interface.
Finally, we conclude our work in \cref{sec:conclusion} and end with
subsequent ideas in \cref{sec:future_work}.

\newpage
\section{Related Work}
\label{sec:related_work}

\subsection{HLS Implementations}
\label{sec:rel_hls_approach}
In the prior work \cite{basu_nist_hls_2019} and the follow-up publication
\cite{soni_hardware_2021}, where the authors expand their results for signature
schemes but report results only for the Artix-7 platform,
\citeauthor{basu_nist_hls_2019} and \citeauthor{soni_hardware_2021}
present the implementation of multiple candidates in round two.
They also list updated results of resource usage, clock, and latency in tables on
their website \cite{soni_nist_nodate} and publish some of their Vivado projects on
Github \cite{soni_deepraj88_nodate}.
We
compare
our designs
with the results of this
work in
\cref{sec:evaluation}.

In \cite{zhao_optimization_2020}, \citeauthor*{zhao_optimization_2020}
extend the \gls{nist} standard interface with a random buffer to be supplied by
the host software application, which is an approach also used in this work.
They also present optimized \gls{hls} designs for the \textit{Kyber512} variant of
CRYSTALS Kyber in round two, which at this point claimed a security Level of 1,
and report an improvement of 74.6\% for the encapsulation and 54.4\% for the
decapsulation compared to the prior work in \cite{basu_nist_hls_2019} 
for the \gls{lap} × clock period metric.
We compare our results with this work in
\cref{sec:evaluation}.
\subsection{HDL Implementations}
\label{sec:rel_hdl_approach}
In \cite{dang_high-speed_2021}, \citeauthor*{dang_high-speed_2021}
evaluate existing hardware implementations for round three \gls{kem} candidates.
They make a distinction between \textit{lightweight} and 
\textit{high-per-formance}
implementations and conclude that \hls implementations are inferior both in
resource usage and performance compared to \gls{hdl} implementations.
They present
new \textit{high-performance} \gls{hdl} designs for CRYSTALS Kyber, Saber, and NTRU
\gls{kem} schemes, which they claim to
be the fastest concerning latency to date.
We compare our results with this work in
\cref{sec:evaluation}.

In \cite{land_hard_2021}, \citeauthor*{land_hard_2021}
present new \textit{lightweight} \fpga implementations of the CRYSTALS Dilithium
\gls{dsa} scheme, of which they claim to make the most efficient use of resources,
surpassing the results of \cite{ricci_implementing_2021}.
They also publish their implementations as open-source, which makes
them a promising \gls{dsa} candidate for integration in \tapasco.
The authors of another \textit{high-performance} Dilithium \gls{hdl} implementation
\cite{beckwith_high-performance_2021} claim to beat this work in
terms of latency but require more resources, which is the reason why we choose
this work for comparison with our results in
\cref{sec:evaluation}.
\subsection{TaPaSCo-related Work}
\label{sec:tapasco_riscv}
In \cite{heinz_catalog_2019}, \citeauthor*{heinz_catalog_2019} present
multiple open-source \riscv cores made ready to use as \tapasco \pes. They
evaluate the performance of their designs in hardware utilizing the same Linux
kernel driver and runtime libraries as this work. This approach serves as an example, which we are
adapting for \pqc accelerators.
These \riscv \pes could be extended with in-pipeline \pqc accelerators leveraging
\gls{hdl} methodologies as in \cite{fritzmann_risq_2020}.
On the contrary, communication with other standalone accelerators as presented
in this work is not implicitly supported, but requires to implement an individual \tapasco interconnect feature.

\section{Approach and Implementation}
\label{sec:approach}

In this section, we integrate \pqc accelerators in
\tapasco. First, we discuss some preliminary considerations.
Then, we take the
ANSI~C reference implementation of CRYSTALS Dilithium and CRYSTALS Kyber as a basis for \hls.
Subsequently, we modify the
\pe interface for Kyber to take a random buffer supplied by the host runtime.
Alongside,
we outline the key points in
the design of the \tapasco host runtime application, which is developed
to test the designs through the \nist \kat mechanism.

Utilizing the \tapasco framework means the functionality of our accelerators
is implemented in \acrfullpl{pe}. A \pe is a hardware kernel containing the
logic to solve a specific task. Usually a \tapasco design is composed of
multiple \pes, which are accompanied by generic components that are necessary to
control their usage such as a \gls{dma} engine, an interrupt controller, and a status core, which stores meta-information about the design that is required by the Linux driver and runtime applications. In this work, we map each operation of a \pqc scheme, such as \textit{Encapsulation} or \textit{Sign}, to one
\pe.
First, we need to define an interface to the \pes that minimizes the amount of
data necessary to start the \pe execution in order to overcome the memory wall.
The communication overhead is implementation-independent but an off-chip
accelerator should unite most functionality to reduce frequent transfers
from and to the general-purpose processor.
Most performance evaluations neglect to measure the impact of the
communication overhead for off-chip accelerators in a wall clock execution time
comparison with the
software
implementation.
For our \tapasco \pes, we 
measure the time necessary to move data from
and to the \pe and compare it with the optimized \acrshort{avx2}
\acrshort{cpu}
implementation to set
the bottom line for a hardware accelerator.
We decide to use the standard \nist interface consisting of
\texttt{crypto\_kem\_enc},
\texttt{crypto\_kem\_dec} for \glspl{kem} and
\texttt{crypto\_sign},
\texttt{crypto\_sign\_open} for
\glspl{dsa}.
This interface has the advantage concerning the memory communication overhead
that the parameters, such as keys and ciphertexts, are stored in compressed form, which
means it is the smallest data to transfer.
Another advantage concerning the host runtime is that all candidates of the \npp are
conforming to this interface, only with different parameter sets. The wrappers around
\texttt{libtapasco} can leverage Rust's \textit{const generics} feature to be generic over the
actually used key sizes. This means if a parameter, like key size, is altered, the
change in the runtime implementation is trivial, and extending the runtime application for
other parameter sets of other candidates is straightforward.
In this work, we use \tapasco's \gls{hls} interface in the version
of commit \texttt{1b88fbc} from the \texttt{develop} branch,
which in turn utilizes the \textit{Xilinx Vitis} toolchain in version \texttt{2022.2}.
With the use of \hls, we focus on larger high-performance \fpgas because \hls from the software reference implementations
is typically inadequate for smaller designs that can fit on \acrshort{iot}-class \fpga devices \cite[4]{dang_high-speed_2021}.
The \fpga devices used throughout this work are the \textit{Xilinx Virtex-7 VC709}
and \textit{Xilinx Alveo U280}.
The current ANSI~C reference implementation of Dilithium \cite[commit
\texttt{3e9b9f1}]{repo_dilithium_2021}
is extended with a \tapasco \texttt{kernel.json} for each of the top-level functions
\texttt{crypto\_sign} and \texttt{crypto\_sign\_open} on every security level.
We
exclude the keypair generation due to its dependence on a
secure \gls{rng}, which is unavailable on the \fpga.
The
respective files
are located in the \texttt{crystals-dilithium} repository in the \texttt{kernel} folder %
(\url{https://github.com/esa-tu-darmstadt/PQC-HA-CRYSTALS-Dilithium/blob/master/kernel/dilithium2_sign/kernel.json}).
Apart from meta information such as name, description, and ID, which should be adapted accordingly, only the \texttt{CompilerFlags} parameter \texttt{DILITHIUM\_MODE} is necessary to change the security level.
Moving
on to CRYSTALS Kyber \cite{bos_kyber_2017}, %
the
software reference implementation \cite[commit \texttt{518de24}]{repo_kyber_2021}
is extended with a \tapasco \texttt{kernel.json} for each of the top-level functions
\texttt{crypto\_kem\_enc} and \texttt{crypto\_kem\_dec} on every security level.
The respective files
are located in the \texttt{crystals-kyber} repository in the \texttt{kernel} folder
(\url{https://github.com/esa-tu-darmstadt/PQC-HA-CRYSTALS-Kyber/blob/master/kernel/kyber2_enc/kernel.json}).
Note, that the Kyber \pes have the number of
the \texttt{KYBER\_K} parameter in their name, which has the values 2, 3 and 4 but
corresponds to \nist security Levels 1, 3, and 5.
As for Dilithium,
only the
\texttt{CompilerFlags} parameter \texttt{KYBER\_K} is necessary to change the security level.
It requires only one modification to
disable the call to the \texttt{randombytes} function for successful \hls compilation.
This function fills a buffer with bytes from the Linux \gls{rng},
which is necessary to create a secure shared secret and is not available on an
\fpga.
At this point, the
Kyber \pes can generate ciphertexts from which the correct shared secret
can be recovered successfully but the buffer usually filled with secure random
data in software is left to consist of arbitrary uninitialized memory.
Therefore, the random buffer used on the \fpga
is initialized from the random data on the host by using the same approach as \cite{zhao_optimization_2020}.
The function \texttt{crypto\_kem\_enc} is modified to receive
its buffer of random data via a new parameter, thereby moving its declaration
into the function signature, while the call to \texttt{randombytes} is still disabled.
This change is guarded by the \texttt{\_\_SYNTHESIS\_\_} conditional compilation
mechanism.
The runtime %
expands
the \texttt{seed} given by the \kat file
in the same manner as the official \nist implementation to give the modified
\textit{Encapsulation} \pe the correctly initialized buffer of random data. With this
modification in place, the Kyber \textit{Encapsulation} \pe produces the same shared secret
and ciphertext as the software implementation and all Kyber \pes are fully verified 
in hardware.
This
verification is achieved with the use of \kat response files (suffixed by \texttt{.rsp}), generated from the software
reference implementation.
These can be found directly
in the \texttt{tapasco-pqc-runtime} repository
(\url{https://github.com/esa-tu-darmstadt/PQC-HA-TaPaSCo-Runtime/blob/main/PQCkemKAT_1632.rsp}).
The runtime first parses the appropriate file for the given algorithm and
security level to retrieve a list of usually \num{100} test cases.
The following procedure is the same for both \glspl{kem} and \glspl{dsa} %
and generic over specific \glspl{kem} and \glspl{dsa}, to be easily extendable
when further
algorithms will be selected by \nist.
The runtime passes the arguments from the test case to the \pe associated with the \texttt{apply}
operation and receives its output, which is passed on to the \pe associated with
the \texttt{verify} operation. The outputs of both \pes are compared with the
respective entries in the \kat case and if all entries match, the test is passed.
This process is repeated for every
test case in the \kat file.
\section{Evaluation}
\label{sec:evaluation}

In this section, we
evaluate the results of the \tapasco \gls{dse} in \cref{sec:dse_results}. Then, we compare our results with related
work using \hls
and for an \gls{hdl} methodology in
\cref{sec:eval_rel_work}.
Finally, we contrast
this with a software implementation
and analyze the real-world communication overhead from the \tapasco host runtime application to the \pes in
\cref{sec:eval_software}.
We implement a bitstream for each \pe of Kyber \textit{Encapsulation} and \textit{Decapsulation} and
Dilithium \textit{Sign} and \textit{Verify} for all security levels
on \textit{Xilinx \vc} and \textit{Alveo \au} \fpgas.
All \pes produce correct results, which are verified by the \tapasco host runtime application.
To measure execution time, the \tapasco host runtime application is invoked \num{1000} times for each \pe and software implementation. The \vc and \au \fpgas are fully connected over \textit{\acrshort{pcie} 3.0} %
and the software runs on an
\textit{AMD EPYC 7443P}
\acrshort{cpu}.
The average of \num{1000} runs is taken of all runs as in \cite{land_hard_2021}.
We
report resource utilization of the \tapasco \textit{user logic}. This
excludes resources required by generic components of every \tapasco design, such
as the \gls{dma} engine, interrupt controller, and status core.

\subsection{TaPaSCo DSE Results}
\label{sec:dse_results}

\begin{figure}[htbp]%
    \centering
    \includegraphics[width=\textwidth]{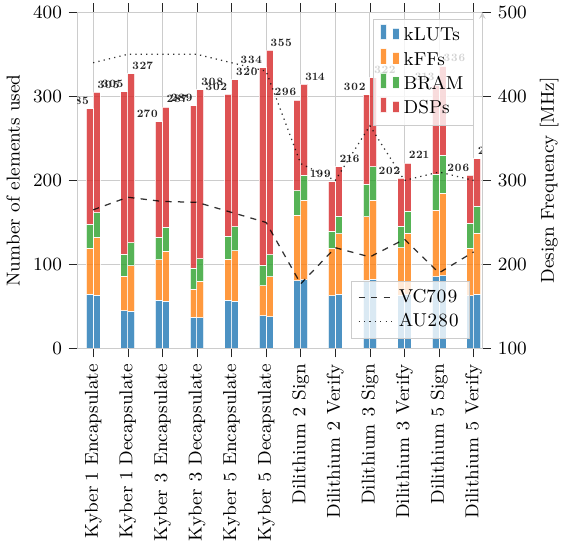}
    \caption{PE Overview of Resource Usage vs. Design Frequency on U280 (left bars) and VC709 (right bars)}
    \label{fig:pe_resource_overview_both_platforms}
\end{figure}

This section describes the results of the \tapasco \gls{dse}.
In \cref{fig:pe_resource_overview_both_platforms}, we plot the resource usage and
design frequency of each \gls{pe} on the \vc and \au
platforms. From left to right each \pe is shown as a triplet of an algorithm, security
level, and operation. The left half displays Kyber, and the right half Dilithium,
while security levels are increasing to the right and operations are
alternating. The left y-axis shows resource usage as the number of elements, where
each color represents another type. \dsps are given in red, \brams in green,
\ffs are given as a multiple of thousand in orange, and \luts also as a multiple of
thousand in blue. The bold black number on top of each bar shows the rounded total sum
of the elements used, where \luts and \ffs enter the calculation as thousands. The left bar of each \pe represents resource usage on the \au. The right bar shows resource usage on the \vc platform. \brams are given as \texttt{RAMB36}
equivalents, and \dsps as \texttt{DSP48} for both platforms. The right
y-axis shows the design frequency in MHz with a dotted line for \au
and a dashed line for \vc.
The resource usage trend across algorithms and operations is identical for both platforms. For both, the number of \luts, \brams, and \dsps is roughly the same, while the number of \ffs is consistently higher on \vc. Concerning algorithms, we can see that Kyber uses fewer \luts, \ffs and \brams in general, than Dilithium, but required more \dsps. An interesting finding is that, while for Kyber with increasing security level usage of \luts, \ffs and \brams declines, the use of \dsps increases, with Kyber 5 \textit{Decapsulation} having the highest usage of \dsps of all. For Dilithium resource usage increases moderately with the security level. With the exception of \dsps of Kyber 5, \textit{Decapsulation} has less resource usage than \textit{Encapsulation} and \textit{Verify} less than \textit{Sign} for Dilithium respectively.
Besides using less resources, the \au reaches almost 1.6 times the design frequency for Kyber and approximately 1.3 times for Dilithium compared to \vc. Kyber achieves higher design frequencies with low variability between
different security levels and operations, while for Dilithium on \vc a clear
distinction between \textit{Sign} and \textit{Verify} \pes is visible. On \vc \textit{Verify} achieves
higher frequencies, but on \au it is \textit{Sign}.
In general, we can see that the \gls{dse} can yield optima of unexpectedly
high frequencies, such as Dilithium 3 \textit{Sign} on \au, but also those lower
than usual, that is Dilithium 2 \textit{Sign} on \vc.
By utilizing the portability of \tapasco designs, moving our \pes to a
newer \fpga model can bring less resource usage while achieving higher design
frequencies.

\begin{figure}[htbp]%
        \centering
        \includegraphics[width=\textwidth]{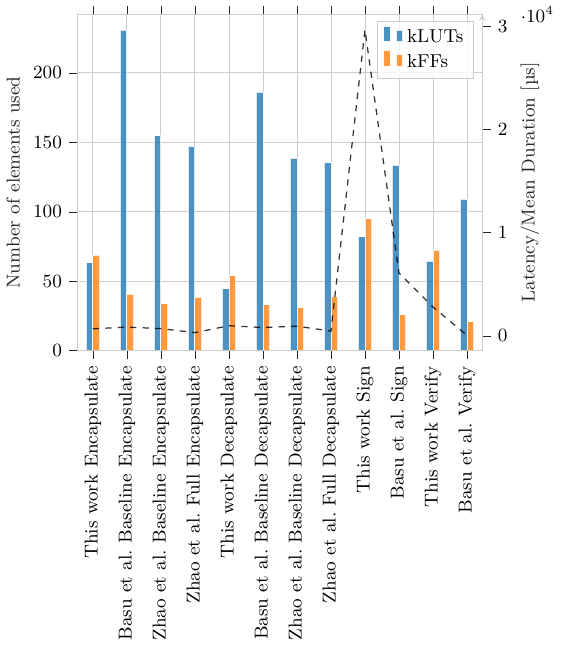}
        \caption{Comparison of TaPaSCo Kyber and Dilithium PEs on Virtex-7 and Security Level 1 with related work using an HLS approach}
        \label{fig:comparison_related_work_kyber1_hls}
\end{figure}

\subsection{Comparison with Related Work}
\label{sec:eval_rel_work}
In this section, we compare our \tapasco implementations with the results of
related work. We first compare the \pes of one scheme with related
work that also uses \hls but reports only \luts and \ffs for
security Level 1.
Then, we compare the \pes of the same scheme with related work that uses an
\gls{hdl} methodology across all security levels.
We use the \gls{dse} results for \vc because related work also reports results
for Virtex-7 or Artix-7 platforms, whereas the \au belongs to another generation
of \fpgas.
We do not apply the \acrfull{lap} as our primary metric because there is no clear
methodology how to weight different elements, that is \ffs are usually cheaper than
\dsps. Therefore we consider resources and latency side-by-side. Additionally, it is unclear
how the latency reports are generated in related work. It looks like the
hardware execution time is calculated by multiplying latency in the average case
with the design frequency, thus not including software overhead. Unfortunately,
Vivado does not report the estimated latency of our \pes, which can nevertheless
vary a lot due to the algorithmic design of the schemes. The average latency can
be analyzed through simulation but for large \hls designs this is slow and for
the amount of our \pes therefore infeasible. Instead, we evaluate average
execution times in the next section from the software standpoint with the
\tapasco runtime by measuring the execution time of our \pes including
interrupt handling and memory communication overhead. %

\begin{figure}[htbp]%
    \centering
        \includegraphics[width=\textwidth]{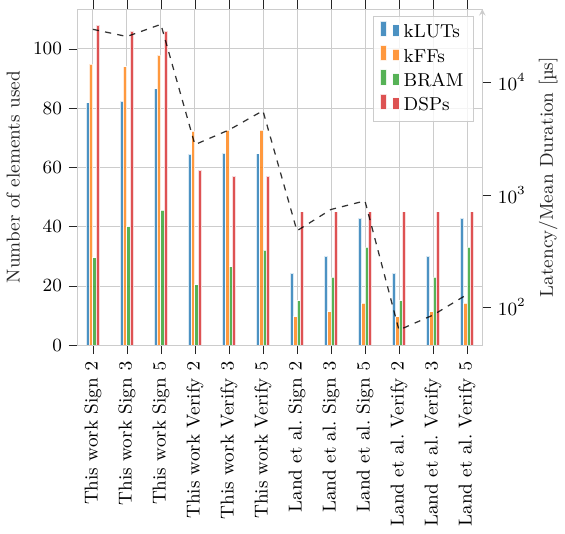}
        \caption{Comparison of TaPaSCo Dilithium PEs on Virtex-7 and related work from Land et al. (HDL) on Artix-7}
        \label{fig:comparison_related_work_dilithium_hdl_log}
\end{figure}

In \cref{fig:comparison_related_work_kyber1_hls}, we compare our Kyber and Dilithium \pes
of security Level 1 on \vc with the results of related work in
\cite{basu_nist_hls_2019} on VC707 and \cite{zhao_optimization_2020} as well as \cite{basu_nist_hls_2019} on \vc.
We only compare \luts and \ffs because the authors of related work do not
report their utilization of \brams and \dsps.
From left to right, each implementation is displayed as a combination of
work/author, optimization level, and operation, where \textit{Encapsulation} is compared
in the
first third,
\textit{Decapsulation} in the
second third
and \textit{Sign} and \textit{Verify} in the last third.
The left y-axis shows
the number of elements used, where \luts and \ffs are given as multiple of
thousand in blue and orange respectively. The right y-axis shows the mean
execution time measured in software for this work and calculated from latency
and frequency for related work in micro-seconds as a dashed line.
For Kyber, we observe that our designs use the least \luts and the most \ffs for both
operations, with the lowest resource utilization in total. The reduced usage of
\luts probably incurs the cost of more \dsps but related work does not report
numbers for their use of this element. However, while the
latency for \textit{Encapsulation} is at the same level as the baseline implementation
from \citeauthor{zhao_optimization_2020} and better than the baseline
implementation of \citeauthor{basu_nist_hls_2019}, the latency for \textit{Decapsulation}
is higher than those of all other implementations. However, their latency report
does not include overhead contained in the result for \tapasco \pes.
Employing the same optimizations as \citeauthor{zhao_optimization_2020} would be
interesting but is not promising when compared to related work using an \gls{hdl}
approach as seen in
\cref{fig:comparison_related_work_dilithium_hdl_log}.

For Dilithium,
we observe that our \pes use fewer \luts and more \ffs than the
implementations of \citeauthor{basu_nist_hls_2019} leading to a higher resource
utilization overall.
To conclude, this means that our \hls approach with minimally modified C code is
more suited for Kyber than Dilithium to produce low resource utilization with
comparable latency but in general yields decent results compared to other
\hls-based works.
\begin{figure}[htbp]%
    \centering
        \includegraphics[width=\textwidth]{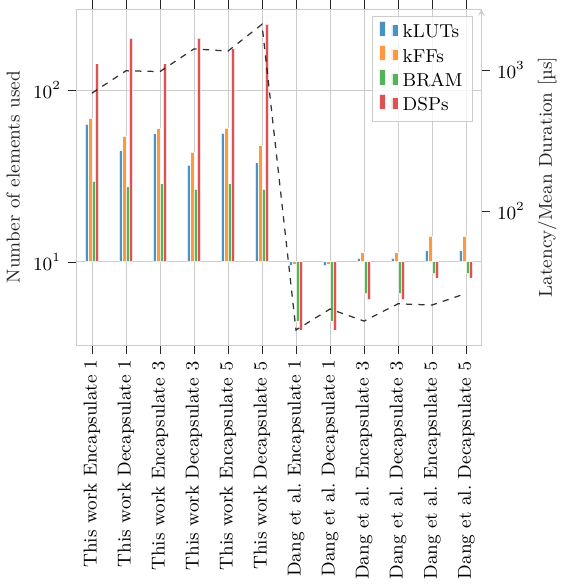}
        \caption{Comparison of TaPaSCo Kyber PEs on Virtex-7 and related work from Dang et al. (HDL) on Artix-7}
        \label{fig:comparison_related_work_kyber_hdl_log}
\end{figure}

%\begin{figure}[htbp]%
%    \centering
%        \includegraphics[width=\textwidth]{externalized/comparison_related_work_dilithium_hdl_log.tikz.pdf}
%        \caption{Comparison of TaPaSCo Dilithium PEs on Virtex-7 and related work from Land et al. (HDL) on Artix-7}
%        \label{fig:comparison_related_work_dilithium_hdl_log}
%\end{figure}

In \cref{fig:comparison_related_work_kyber_hdl_log}, we compare our Kyber \pes
on all security levels with the results of related work in
\cite{dang_high-speed_2021}.
From left to right, each \pe is shown as a triplet of work/author, operation, and
security level. The left half displays our Kyber \pes and the right half the
implementations from \citeauthor{dang_high-speed_2021},
while security levels are increasing to the right and operations are
alternating. The left y-axis shows resource usage as the number of elements in a
logarithmic scale, where
each color represents another type. \luts are given as a multiple of thousand in
blue, \ffs also as a multiple of thousand in orange, \brams in green and \dsps are
given in red. The right y-axis shows the mean execution time measured in
software for this work and calculated from latency and frequency for related
work in micro-seconds as a dashed line in a logarithmic scale.
This diagram affirms the conclusion of \citeauthor{dang_high-speed_2021} that
\hls implementations cannot keep up with \gls{hdl} designs. Our \pes
have roughly a magnitude higher resource utilization, even though they are
specialized for each operation and their core is capable of both operations and
key generation for each security level. Finally, their latency/mean execution
time is also more than a magnitude lower but this still does not yield enough
performance to overcome the memory wall as we show in the next section.

In \cref{fig:comparison_related_work_dilithium_hdl_log}, we compare our Dilithium \pes
on all security levels with the results of related work in
\cite{land_hard_2021}.
From left to right, each \pe is shown as a triplet of work/author, operation, and
security level. The left half displays our Dilithium \pes and the right half the
implementations from \citeauthor{land_hard_2021},
while security levels are increasing to the right and the \textit{Sign} operation is
shown first, Verification second.
The y-axis shows resource usage and mean execution time as in the last diagram
\cref{fig:comparison_related_work_kyber_hdl_log}.
This diagram confirms the conclusion of \citeauthor{dang_high-speed_2021} that
\hls implementations cannot keep up with \gls{hdl} designs for Dilithium too.
Our \pes have roughly a magnitude higher resource utilization, even though they
are specialized for each operation and their core is capable of both operations
and key generation. Finally, their latency/mean execution time is also more than a magnitude lower. Yet, this does not yield enough performance to compete with the optimized \gls{avx2} implementation, as we will see in the next section.

\subsection{Comparison of Execution Time against Software}
\label{sec:eval_software}

\begin{figure}[htbp]%
        \includegraphics[width=0.95\textwidth]{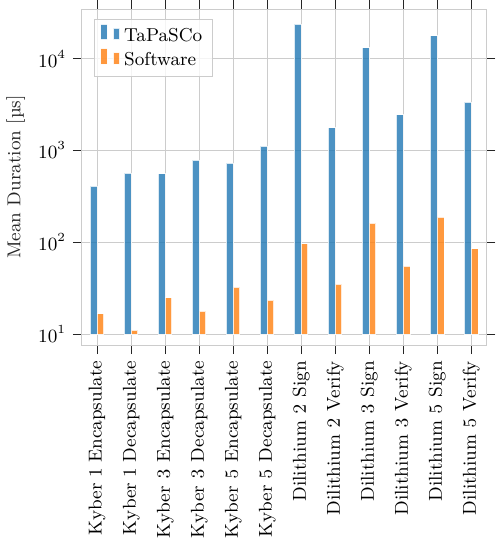}
        \caption{Mean Execution Time of \tapasco on U280 and \gls{avx2} Software Implementation on AMD Epyc}
        \label{fig:mean_duration_software_vs_pes_logy}
\end{figure}

\begin{figure}[htbp]%
        \includegraphics[width=0.95\textwidth]{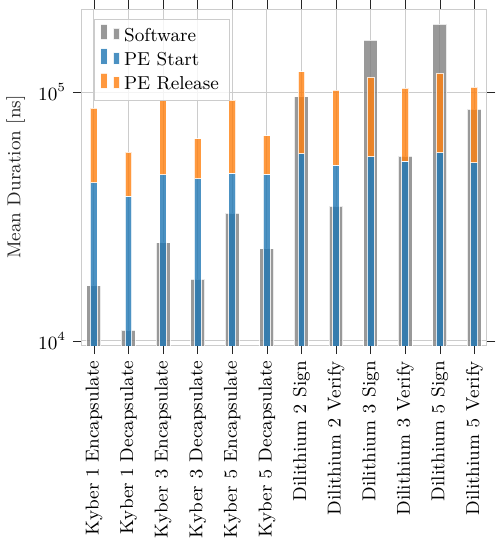}
        \caption{Communication Overhead of \tapasco on U280 and Mean Execution Time in \gls{avx2} Software Implementation on AMD Epyc}
        \label{fig:communication_overhead_pes_vs_software}
\end{figure}

In this section, we compare our results against the software implementations of
Kyber and Dilithium. As \tapasco \pes are standalone accelerators, we evaluate
the communication overhead introduced by the host and \fpga memories connected
through \textit{\gls{pcie} 3.0}%
. We test under realistic conditions from the same application on the same host and also
include software interrupt handling overhead.

In \cref{fig:mean_duration_software_vs_pes_logy}, we compare the execution
time from the \tapasco host runtime application with the optimized \gls{avx2}
implementation running on the host \gls{cpu}. 
From left to right, all our implementations are listed as triplets of the algorithm,
security level, and the operation. The left half shows results for Kyber, the right
half for Dilithium. The y-axis shows the mean execution time in micro-seconds on a
logarithmic scale of both \tapasco in blue bars and the \gls{avx2} software
implementation in orange bars.
We observe that the software implementation is roughly one magnitude faster than
the \tapasco \pes.

In \cref{fig:communication_overhead_pes_vs_software}, we compare the
communication overhead of the \tapasco host runtime application with the
optimized \gls{avx2} Implementation running on the host \gls{cpu}. 
From left to right, all our implementations are listed as a triplet of algorithm,
security level, and operation. The left half shows results for Kyber, the right
half for Dilithium. The y-axis shows the mean execution time in nano-seconds on a
linear scale of the \gls{avx2} software
implementation in gray bars and the communication overhead for preparation of a
\pe as blue and the release of a \pe as orange bar stacked on top of each other.
Across all \pes, the communication overhead is around/less than \SI{10000}{\nano\s}.
Mean execution time in software increases generally with security level, while
for Kyber both operations have roughly the same value with \textit{Decapsulation} only
slightly lower than \textit{Encapsulation}, Dilithium varies greatly between signing and
verification.
We observe that for Kyber 1 \textit{Encapsulation} the communication overhead alone is
almost as high as the effort to solve the original problem itself. For Kyber 1
\textit{Decapsulation} the communication overhead alone is already higher than the
software implementation. For Kyber 3 the overhead takes up half of the software
implementation and for Kyber 5 approximately a third.
We conclude that using the \tapasco architecture accelerating Kyber on security
Level 1 is impossible on the evaluated platform and even using the
state-of-the-art \gls{hdl} accelerator from \citeauthor{dang_high-speed_2021}
would not be faster than the \gls{avx2} software implementation. The same holds
for the Dilithium implementation from \citeauthor{land_hard_2021}, assuming
the same communication overhead for their interface when integrated into the
\tapasco architecture.
However, this only holds true if the only metric employed is latency. The
obvious next step to evaluate throughput on \textit{high-performance} \fpga devices is
to look at \textit{pipelining}. Other scaling mechanisms such as using multiple \pes
concurrently in a cluster are also possible and can be combined with multi-potent
\gls{hdl} designs, where each \pe can perform all operations of the
scheme, \eg key generation, encapsulation, and decapsulation for a
\gls{kem}, and executes the desired operation according to the current demand.
Accumulating multiple outstanding memory requests could also result in a benefit
concerning the communication overhead and lower overhead could in turn increase
throughput again.
Finally, regarding these results for high-performance standalone accelerators, it would be
interesting to see more in-pipeline accelerators like those of
\cite{fritzmann_risq_2020} but for \textit{high-performance} \acrshort{cpu}s.

\section{Conclusion}
\label{sec:conclusion}

We integrated the software implementations of two algorithms from the \npp into a generic \fpga runtime utilizing the features of \tapasco via \gls{hls}. We present a bitstream for CRYSTALS Kyber and Dilithium on all
three security levels for \textit{Encapsulation}/\textit{Sign} and
\textit{Decapsulation}/\textit{Verify}. Our approach leads to designs with
comparable resource utilization and performance to baseline results of
related work also using \hls while achieving a higher design
frequency as a result of the \tapasco \dse.
They are verified in hardware to produce the same correct
results as the software implementation through the \kat mechanism.
This straightforward methodology is adequate to develop a \tapasco host runtime
application and evaluate the \pe interface including the corresponding
communication overhead but is clearly inferior in terms of resource utilization
and performance compared to \gls{hdl} designs of related work.
We demonstrate portability by synthesizing
the same designs for \vc and \au, thereby
gaining faster designs while lowering resource utilization.
We analyze the communication overhead for standalone memory-mapped accelerators
and conclude that by utilizing the \tapasco \pe architecture, there is only a tight
window for speeding up Kyber against the optimized reference \gls{avx2}
implementation. For Kyber on security Level 1, the communication overhead is even as high
as the time necessary to solve the operation in software. The window for
Dilithium is larger, especially for the Sign operation, but assuming the same
overhead, it is questionable if even state-of-the-art \gls{hdl} implementations
could beat the optimized software implementation in the high-performance domain.
To wrap up, \tapasco can aid with the design process by using its \pe architecture and
shines in testing of \pqc accelerators on real hardware utilizing its Linux kernel
driver and our generic runtime. For the \npp specifically, the ANSI C reference
implementations are straightforward to integrate into the \tapasco ecosystem via
\hls but this does not replace the \gls{hdl} methodology.

\section{Future Work}
\label{sec:future_work}

With the foundation for \tapasco \pqc accelerators laid in this work, there are
multiple promising ways for further exploration.
First,
existing hardware implementations that are published as open-source can be
integrated in the \tapasco \pe architecture, where lightweight designs open the
possibility to make use of the diverse \tapasco device support for
smaller \fpgas. There, the communication overhead can be re-evaluated on
embedded platforms that use a different memory architecture, such as shared memory
on the Ultra96%
, and compared to the optimized software implementations of
other \gls{cpu} \glspl{isa}, \eg
Neon \acrshort{ntt} \cite{becker_neon_2021}
on ARM.
Finally, the \hls-based \tapasco approach as shown in this work can be applied
to other, more unexplored, alternate candidates of the \npp for the next round,
where all \tapasco components, including the host runtime application can be
re-used because all candidates are required to conform to the same interface.

\printbibliography{}
\begin{landscape}
\begin{table}[p]
\tiny
\caption{List of all \pes with design frequency and resource utilization}
\label{tab:tapasco_pe_resource_frequency_results}
\begin{tabular}{lllSlllSSSSS}
\toprule
{} & {Algorithm} & {Platform} & {Security Level} & {Operation} & {PE Name} & {Functionality} & {Frequency [MHz]} & {kLUTs} & {kFFs} & {BRAM} & {DSPs} \\
\midrule
0 & Kyber & VC709 & 1 & Encapsulate & kyber2\_enc & Working & 265 & 63.329000 & 68.883000 & 29.500000 & 143.000000 \\
1 & Kyber & VC709 & 1 & Decapsulate & kyber2\_dec & Working & 280 & 44.488000 & 53.918000 & 27.500000 & 201.000000 \\
2 & Kyber & VC709 & 3 & Encapsulate & kyber3\_enc & Working & 275 & 55.982000 & 59.783000 & 28.500000 & 143.000000 \\
3 & Kyber & VC709 & 3 & Decapsulate & kyber3\_dec & Working & 274 & 36.704000 & 43.616000 & 26.500000 & 201.000000 \\
4 & Kyber & VC709 & 5 & Encapsulate & kyber4\_enc & Working & 262 & 56.166000 & 60.297000 & 28.500000 & 175.000000 \\
5 & Kyber & VC709 & 5 & Decapsulate & kyber4\_dec & Working & 250 & 37.864000 & 47.445000 & 26.500000 & 243.000000 \\
6 & Kyber & AU280 & 1 & Encapsulate & kyber2\_enc & Working & 440 & 64.900000 & 54.792000 & 28.500000 & 137.000000 \\
7 & Kyber & AU280 & 1 & Decapsulate & kyber2\_dec & Working & 450 & 45.038000 & 40.858000 & 26.500000 & 193.000000 \\
8 & Kyber & AU280 & 3 & Encapsulate & kyber3\_enc & Working & 450 & 57.164000 & 48.447000 & 27.000000 & 137.000000 \\
9 & Kyber & AU280 & 3 & Decapsulate & kyber3\_dec & Working & 450 & 37.307000 & 33.600000 & 25.000000 & 193.000000 \\
10 & Kyber & AU280 & 5 & Encapsulate & kyber4\_enc & Working & 440 & 57.745000 & 48.585000 & 27.000000 & 169.000000 \\
11 & Kyber & AU280 & 5 & Decapsulate & kyber4\_dec & Working & 430 & 39.127000 & 35.664000 & 24.500000 & 235.000000 \\
12 & Dilithium & VC709 & 2 & Sign & dilithium2\_sign & Deadlock & 177 & 81.877000 & 94.875000 & 29.500000 & 108.000000 \\
13 & Dilithium & VC709 & 2 & Verify & dilithium2\_verify & Working & 220 & 64.537000 & 72.207000 & 20.500000 & 59.000000 \\
14 & Dilithium & VC709 & 3 & Sign & dilithium3\_sign & Working & 209 & 82.336000 & 94.028000 & 40.000000 & 106.000000 \\
15 & Dilithium & VC709 & 3 & Verify & dilithium3\_verify & Working & 230 & 64.691000 & 72.448000 & 26.500000 & 57.000000 \\
16 & Dilithium & VC709 & 5 & Sign & dilithium5\_sign & Working & 190 & 86.601000 & 97.852000 & 45.500000 & 106.000000 \\
17 & Dilithium & VC709 & 5 & Verify & dilithium5\_verify & Working & 215 & 64.569000 & 72.445000 & 32.000000 & 57.000000 \\
18 & Dilithium & AU280 & 2 & Sign & dilithium2\_sign & Deadlock & 320 & 80.943000 & 77.604000 & 29.000000 & 108.000000 \\
19 & Dilithium & AU280 & 2 & Verify & dilithium2\_verify & Working & 300 & 62.922000 & 56.764000 & 20.000000 & 59.000000 \\
20 & Dilithium & AU280 & 3 & Sign & dilithium3\_sign & Working & 365 & 81.262000 & 75.504000 & 39.000000 & 106.000000 \\
21 & Dilithium & AU280 & 3 & Verify & dilithium3\_verify & Working & 300 & 63.204000 & 56.795000 & 25.500000 & 57.000000 \\
22 & Dilithium & AU280 & 5 & Sign & dilithium5\_sign & Working & 310 & 85.421000 & 79.031000 & 43.000000 & 106.000000 \\
23 & Dilithium & AU280 & 5 & Verify & dilithium5\_verify & Working & 300 & 62.857000 & 56.795000 & 29.500000 & 57.000000 \\
\bottomrule
\end{tabular}
\end{table}

\begin{table}
\tiny
\caption{List of all \pes with mean execution time and communication overhead}
\label{tab:tapasco_pe_duration_overhead_results}
\begin{tabular}{lllSlllSSSSS}
\toprule
{} & {Algorithm} & {Platform} & {Security Level} & {Operation} & {PE Name} & {Functionality} & {Frequency [MHz]} & {PE Start [ns]} & {PE Wait [ns]} & {PE Release [ns]} & {Mean Duration [$\mu$s]} \\
\midrule
0 & Kyber & VC709 & 1 & Encapsulate & kyber2\_enc & Working & 265 & 33879.021000 & 625722.209000 & 24261.589000 & 683.862819 \\
1 & Kyber & VC709 & 1 & Decapsulate & kyber2\_dec & Working & 280 & 38860.905000 & 927236.989000 & 17254.458000 & 983.352352 \\
2 & Kyber & VC709 & 3 & Encapsulate & kyber3\_enc & Working & 275 & 39136.417000 & 908713.685000 & 23993.971000 & 971.844073 \\
3 & Kyber & VC709 & 3 & Decapsulate & kyber3\_dec & Working & 274 & 36845.543000 & 1353062.283000 & 14922.066000 & 1404.829892 \\
4 & Kyber & VC709 & 5 & Encapsulate & kyber4\_enc & Working & 262 & 33415.510000 & 1300124.540000 & 21841.251000 & 1355.381301 \\
5 & Kyber & VC709 & 5 & Decapsulate & kyber4\_dec & Working & 250 & 40440.257000 & 2036665.064000 & 16949.798000 & 2094.055119 \\
6 & Kyber & AU280 & 1 & Encapsulate & kyber2\_enc & Working & 440 & 43271.507000 & 320968.749000 & 42768.400000 & 407.008656 \\
7 & Kyber & AU280 & 1 & Decapsulate & kyber2\_dec & Working & 450 & 38101.868000 & 505237.785000 & 19285.876000 & 562.625529 \\
8 & Kyber & AU280 & 3 & Encapsulate & kyber3\_enc & Working & 450 & 46614.493000 & 463630.703000 & 46358.170000 & 556.603366 \\
9 & Kyber & AU280 & 3 & Decapsulate & kyber3\_dec & Working & 450 & 44759.069000 & 717163.692000 & 20182.781000 & 782.105542 \\
10 & Kyber & AU280 & 5 & Encapsulate & kyber4\_enc & Working & 440 & 47224.161000 & 635546.020000 & 45309.562000 & 728.079743 \\
11 & Kyber & AU280 & 5 & Decapsulate & kyber4\_dec & Working & 430 & 46826.185000 & 1036884.439000 & 20067.314000 & 1103.777938 \\
12 & Dilithium & VC709 & 2 & Sign & dilithium2\_sign & Deadlock & 177 & 49764.787000 & 29582405.304000 & 43248.174000 & 29675.418265 \\
13 & Dilithium & VC709 & 2 & Verify & dilithium2\_verify & Working & 220 & 37304.096000 & 2728635.281000 & 29307.401000 & 2795.246778 \\
14 & Dilithium & VC709 & 3 & Sign & dilithium3\_sign & Working & 209 & 45576.437000 & 25517599.169000 & 38890.637000 & 25602.066243 \\
15 & Dilithium & VC709 & 3 & Verify & dilithium3\_verify & Working & 230 & 40676.377000 & 3705498.223000 & 27473.324000 & 3773.647924 \\
16 & Dilithium & VC709 & 5 & Sign & dilithium5\_sign & Working & 190 & 44947.777000 & 32771644.154000 & 42263.676000 & 32858.855607 \\
17 & Dilithium & VC709 & 5 & Verify & dilithium5\_verify & Working & 215 & 40420.255000 & 5532179.863000 & 27110.452000 & 5599.710570 \\
18 & Dilithium & AU280 & 2 & Sign & dilithium2\_sign & Deadlock & 320 & 56567.794000 & 23444002.396000 & 64092.834000 & 23564.663024 \\
19 & Dilithium & AU280 & 2 & Verify & dilithium2\_verify & Working & 300 & 50752.353000 & 1660295.248000 & 51009.036000 & 1762.056637 \\
20 & Dilithium & AU280 & 3 & Sign & dilithium3\_sign & Working & 365 & 55028.980000 & 12932134.169000 & 60011.130000 & 13047.174279 \\
21 & Dilithium & AU280 & 3 & Verify & dilithium3\_verify & Working & 300 & 52756.371000 & 2334493.162000 & 51142.836000 & 2438.392369 \\
22 & Dilithium & AU280 & 5 & Sign & dilithium5\_sign & Working & 310 & 57123.034000 & 17644266.313000 & 61795.863000 & 17763.185210 \\
23 & Dilithium & AU280 & 5 & Verify & dilithium5\_verify & Working & 300 & 52060.355000 & 3220930.649000 & 52414.143000 & 3325.405147 \\
\bottomrule
\end{tabular}
\end{table}

\end{landscape}
\end{document}